\documentclass[english]{paper}
\usepackage{amsmath}
\usepackage{tgtermes}

\usepackage{newtxmath}
\usepackage[LGR,T1]{fontenc}
\usepackage[utf8]{inputenc}
\usepackage{geometry}
\usepackage{textcomp}
\usepackage{mathtools}
\usepackage{graphicx}
\PassOptionsToPackage{version=3}{mhchem}
\usepackage{mhchem}

\makeatletter

\ProvideTextCommand{\~}{LGR}[1]{\char126#1}

\newcommand{\lyxaddress}[1]{
	\par {\raggedright #1
	\vspace{1.4em}
	\noindent\par}
}

\usepackage[unicode=true, bookmarks=true,bookmarksnumbered=false,bookmarksopen=false, breaklinks=false,pdfborder={0 0 0},pdfborderstyle={},backref=false,colorlinks=false]{hyperref}
\hypersetup{pdftitle={Lenseless X-ray Nano-Tomography down to 150nm Resolution: On the Quantification of Modulation Transfer and Focal Spot of the Lab-based ntCT System}, pdfauthor={Jonas Graetz, Dominik Müller, Andreas Balles, Christian Fella}}
\usepackage{scalerel}
\usepackage{tikz}
\usetikzlibrary{svg.path}
\definecolor{orcidlogocol}{HTML}{A6CE39}
\tikzset{
  orcidlogo/.pic={
    \fill[orcidlogocol] svg{M256,128c0,70.7-57.3,128-128,128C57.3,256,0,198.7,0,128C0,57.3,57.3,0,128,0C198.7,0,256,57.3,256,128z};
    \fill[white] svg{M86.3,186.2H70.9V79.1h15.4v48.4V186.2z}
                 svg{M108.9,79.1h41.6c39.6,0,57,28.3,57,53.6c0,27.5-21.5,53.6-56.8,53.6h-41.8V79.1z M124.3,172.4h24.5c34.9,0,42.9-26.5,42.9-39.7c0-21.5-13.7-39.7-43.7-39.7h-23.7V172.4z}
                 svg{M88.7,56.8c0,5.5-4.5,10.1-10.1,10.1c-5.6,0-10.1-4.6-10.1-10.1c0-5.6,4.5-10.1,10.1-10.1C84.2,46.7,88.7,51.3,88.7,56.8z};
  }
}
\newcommand\orcidicon[1]{\href{https://orcid.org/#1}{\mbox{\scalerel*{
\begin{tikzpicture}[yscale=-1,transform shape]
\pic{orcidlogo};
\end{tikzpicture}
}{|}}}}

\usepackage{fancyhdr}
\usepackage[font={small,sf},labelfont=bf,format=plain]{caption}

\makeatother

\usepackage{babel}
\begin{document}
\title{Lenseless X-ray Nano-Tomography down to 150$\,$nm Resolution: On
the Quantification of Modulation Transfer and Focal Spot of the Lab-based
ntCT System}
\author{J. Graetz\,\orcidicon{0000-0002-4403-3686}, D. Müller, A. Balles,
C. Fella}
\maketitle

\lyxaddress{\emph{\footnotesize{}Fraunhofer IIS, division EZRT, department MRB,
Josef-Martin-Weg 63, 97074 Würzburg}\\
\emph{\footnotesize{}Lehrstuhl für Röntgenmikroskopie, Universität
Würzburg, Josef-Martin-Weg 63, 97074 Würzburg, Germany}}
\begin{abstract}
The ntCT nano tomography system is a geometrically magnifying X-ray
microscopy system integrating the recent Excillum NanoTube nano-focus
X-ray source and a CdTe photon counting detector from Dectris. The
system's modulation transfer function (MTF) and corresponding point
spread function (PSF) are characterized by analyzing the contrast
visibility of periodic structures of a star pattern featuring line
width from 150$\,$nm to 1.5$\,$μm. The results, which can be attributed
to the characteristics of the source spot, are crosschecked by scanning
the source's electron focus over an edge of the structured transmission
target in order to obtain an independent measurement of its point
spread function. For frequencies above 1000 linepairs/mm, the MTF
is found to correspond to a Gaussian PSF of 250$\,$nm full width
at half maximum (FWHM). The lower frequency range down to 340 linepairs/mm
shows an additional Gaussian contribution of 1$\,$μm FWHM. The resulting
resolution ranges at 3200 linepairs/mm, which is consistent with the
visual detectability of the smallest 150$\,$nm structures within
the imaged star pattern. \thispagestyle{fancy}
\end{abstract}

\section{Introduction}

3D X-ray microscopy at the laboratory is a valuable tool both for
life sciences \cite{Busse2018,Eckermann2020} and materials research
\cite{Maire2014}, whereby one of the central physical and technical
challenges of X-ray microscopy lies in the focusing of X-rays. While
fundamentally possible, and indeed applied in full-field transmission
X-ray microscopes (cf.\ e.g.\ \cite{Takman2007,Seim2012,Rehbein2012,Oton2015}),
it is generally associated with considerable technical challenges
and compromises particular with respect to the accessible X-ray energies.
The use of X-ray optical elements is thus often avoided by either
magnifying in the optical regime (using optical microscopes coupled
to thin scintillating screens close to the sample) or by better focusing
the X-ray source's electron beam, placing the sample close to the
source and utilizing large geometric magnification factors. A brief
summary of different approaches is e.g.\ given by Withers \cite{Withers2007}.
 X-ray microscopy systems based on highly focusing X-ray sources
and large geometric magnification have seen regular attention over
the past decade, and are particularly interesting for their compatibility
with higher X-ray energies. In the past, respective systems maximizing
the achievable resolution were based on repurposed scanning electron
microscopes with optimized electron focusing and suited target design
to achieve reported resolutions in the $10^{-7}\,\mathrm{m}$ regime
\cite{Mayo2003,Brownlow2006,Hanke2009,Bruyndonckx2010,Nachtrab2011,Stahlhut2016,GomesPerini2017}. 

Within the collaborative nanoXCT project \cite{Nachtrab2014}, a dedicated
nano-focus X-ray source has been developed and made commercially available
by Excillum AB (Kista, Sweden), facilitating routine high resolution
imaging on the hundred nanometer scale. It operates in transmission
geometry, whereby a thin tungsten layer of few hundred nanometers
thickness deposited on a 100$\,$\textmu m thick diamond vacuum window
serves as X-ray target. This source has been integrated into our follow-up
in-house nano tomography system ntCT \cite{Fella2018}, which is further
equipped with high precision positioning stages for tomography applications
and a photon counting detector for the efficient detection of X-rays.
First benchmarks of the system have recently been given in \cite{Zabler2020,Graetz2020}.

In order to provide a quantitative assessment of the system's resolution,
the modulation transfer function (MTF) and point spread function (PSF)
will be quantified here using both a star pattern and an edge scan.
Respective techniques have e.g.\ been proposed for lower resolutions
in photography (cf.\ \cite{Loebich2007,Masaoka2010} and ISO 12233:2014/2017)
and have also been used in the context of X-ray imaging using a variety
of methodologies, including classic contrast analysis \cite{Takman2007,Rehbein2012},
Fourier analysis \cite{Weiss2012,Seim2012,Oton2015}, and oriented
edge spread analysis \cite{Probst2019}. Theoretically, the full
2D complex valued optical transfer function may be deduced from star
patterns \cite{Burgess1977}, facilitating spatial reconstructions
of arbitrarily shaped focal spots as has e.g.\ recently been reported
by \cite{Probst2019}. 

The deduction of MTFs from periodic line patterns using Fourier analysis
is generally expected to be more reliable as compared to common line
spread and edge spread based techniques both due to the reduced dependence
on the shape of the pattern \cite{Loebich2007,Horstmeyer2016} and
the reduced noise susceptibility \cite{Oton2015,Gonzalez-Lopez2018}.
This is particularly relevant in the context of the considered resolution
and respective structure scales, which are not covered by typical
standards on X-ray focal spot characterization (cf.\ e.g.\ \cite{Salamon2008}).
In order to nevertheless independently support the results obtained
from the star pattern analysis, a more direct observation of the source's
focal spot is additionally obtained by scanning the focused electron
beam over an edge of the source's structured target layer.

\section{Methods}

\subsection{Experiment}

The experimental setup comprises an Excillum NanoTube N2 nano-focus
X-ray source operated at 60keV, a CdTe-based Dectris Säntis photon
counting detector with a pixel pitch of 75$\,$\textmu m (Dectris
Ltd., Baden-Daettwil, Switzerland), precision positioning actuators,
and a lithographically etched Siemens star made by ZonePlates Ltd.\
(London, United Kingdom) in a tungsten substrate of 1.5$\,$\textmu m
thickness featuring 64 spokes ranging between 150$\,$nm and 1.5$\,$\textmu m
linewidth. The detector is placed at 0.3$\,$m distance from the focal
point. The Siemens star is placed directly in front of the source
yielding a magnification factor of approx.\ $10^{3}$, such that
the smallest structures are well sampled on the detector at an effective
pixel size of 71$\,$nm, while still covering a reasonable field of
view. The almost pixel-confined point spread of the photon counting
detector (cf.\ \cite{Donath2013}) thereby ensures that observed
visibility reductions within the given star pattern can be fully attributed
to the X-ray source.  

Images are normalized (by division) to reference images acquired without
sample. Explicit corrections for dark current are not required due
to the working principle of photon counting detectors, which suppress
thermal noise by means of energy thresholding \cite{Henrich2009}.
In order to improve signal to noise ratio and sampling, 15 acquisitions
of 20 seconds exposure time each are combined. The images are upscaled
by a factor of four and corrected for sub-pixel shifts (cf.\ e.g.\
Dreier et al.\ \cite{Dreier2020} for a more detailed description)
using linear interpolation prior to averaging, yielding a virtual
sampling pitch of 17.8$\,$nm. This approach is analog to the use
of slanted edges (as opposed to edges aligned with the detector grid)
in order to increase the sampling of edge spread functions (an example
of slanted edge analysis can e.g.\ be found in \cite{Donath2013}). 

The employed X-ray source further allows to control the electron beam
deflection, allowing to scan the beam over edges of the source's structured
transmission target.  By measuring the change in X-ray emission as
a function of motion distance perpendicular to the considered edge,
the point spread of the electron beam can be estimated.  In order
to improve the sampling of the edge spread function, which is restricted
by the resolution of deflection current control, the effective motion
component perpendicular to the edge is downscaled by means of scanning
at shallow angles (in analogy to the common slanted edge approach).
The derivative of the so acquired edge spread function corresponds,
under the assumption of negligible influence of the diamond substrate,
perfect edge and exact scanning motion, to the X-ray focal spot's
point spread function.

\begin{figure}

\centering{}\includegraphics[width=0.95\textwidth]{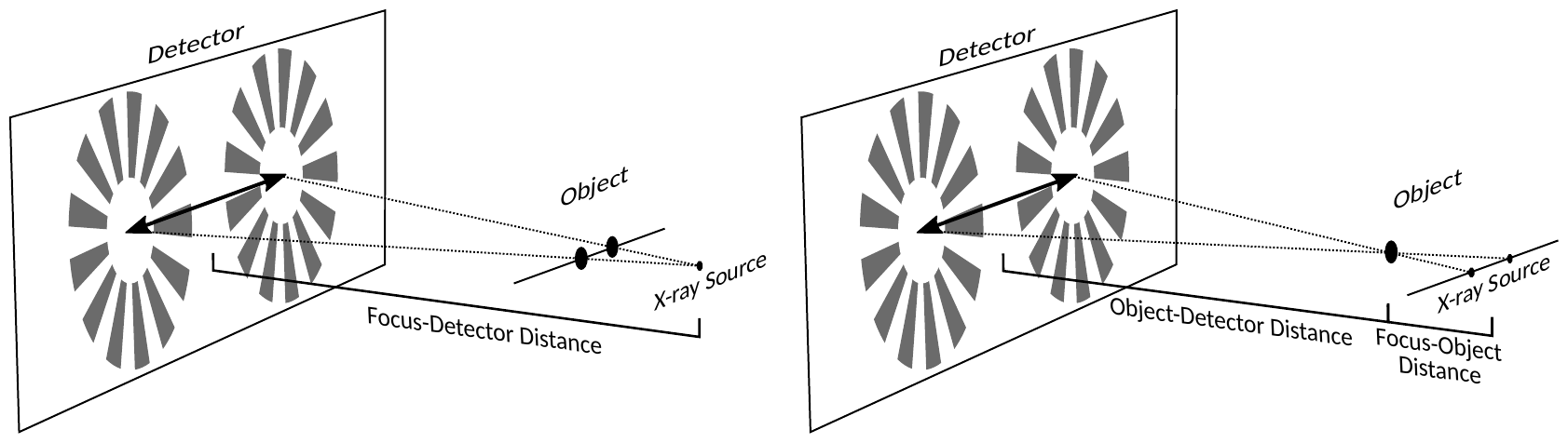}\caption{\label{fig:spatial-calibration}Determination of magnification factor
and focus-object distance by means precisely defined object translation
and known detector distance and extent (left sketch). Based on the
so determined focus-object distance, also the focal spot motion (controlled
by variation of deflection coil currents) is calibrated in spatial
units (right sketch).}
\end{figure}

The spatial dimensions of the star pattern as well as the electron
spot motion are calibrated by means of translating the object and
the focal spot respectively and observing the projected translation
on the detector, as depicted in Figure~\ref{fig:spatial-calibration}.
The known detector pixel size, source--detector distance and translation
distances of the motion controllers thereby serve as references.

\subsection{Siemens star contrast visibility analysis\label{subsec:siemens-star-analysis}}

The image of the Siemens star is transformed to polar coordinates,
yielding an intensity pattern
\begin{equation}
I(r,\varphi)\:,
\end{equation}
with $r\in[0,30\text{\textmu m]}$ and $\varphi\in[0,2\pi[$. In this
representation, the Siemens star resembles a linear line pattern oriented
along the radial dimension and periodic along the angular coordinate.
The real space period length $T$ of the pattern varies along the
radial dimension and is given by
\begin{equation}
\hphantom{f(r)}\mathllap{T(r)}=\mathrlap{\frac{2\pi r}{N}\,,}\hphantom{\frac{1}{T(r)}=\frac{1}{r}\frac{N}{2\pi}\,.}\label{eq:pattern-period}
\end{equation}
with $N$ denoting the number of line pairs within the siemens star,
and $2\pi r$ being its total circumference at a given radius $r$.
Here,
\begin{equation}
\hphantom{f(r)}\mathllap{N}=\mathrlap{64\:.}\hphantom{\frac{1}{T(r)}=\frac{1}{r}\frac{N}{2\pi}\,.}
\end{equation}
The spatial frequency in terms of line pairs per unit length is given
by the inverse period length:
\begin{equation}
f(r)=\frac{1}{T(r)}=\frac{1}{r}\frac{N}{2\pi}\,.\label{eq:pattern-frequency}
\end{equation}

Variations in contrast visibility along the radial dimension, and
thus in dependence of the spatial pattern frequency, provide direct
information on the effective modulation transfer function (MTF). Provided
that the machining precision of the Siemens star is sufficiently higher
than the expected performance of the imaging system, i.e., provided
that the actual profile of the Siemens star can be considered constant
along the radial dimension, the resulting MTF can be directly associated
to the imaging system.

Potential spatial variations in substrate thickness, illumination
or actual contrast visibility as well as missing image regions are
considered by partitioning the Siemens star into 
\begin{equation}
M=8
\end{equation}
segments of 45° covering $N/M=8$ linepairs each, and performing the
analysis independently for each angular segment. 

The contrast visibility $v$ is quantified by means of Fourier analysis:
\begin{align}
v^{(k)}(r) & =\frac{a_{1}^{(k)}(r)}{t^{(k)}(r)}\,,\label{eq:visibility}\\
a_{1}^{(k)}(r) & =\Biggl|\frac{M}{\pi}\int_{k2\pi/M}^{(k+1)2\pi/M}I(r,\varphi)e^{iN\varphi}\,\mathrm{d}\varphi\Biggr|\label{eq:amplitude}\\
t^{(k)}(r) & =\,\frac{M}{2\pi}\int_{k2\pi/M}^{(k+1)2\pi/M}I(r,\varphi)\,\mathrm{d}\varphi\label{eq:mean-offset}
\end{align}
with $(k)$ enumerating the angular sections, $a_{1}^{(k)}(r)$ denoting
the amplitude of the first harmonic of the periodic pattern within
the respective section and $t^{(k)}(r)$ denoting its mean value (i.e.,
the mean X-ray transmission). The ratio $N/M$ must be integer to
ensure that complete periods are covered. The analysis is, due to
the orthogonality of the Fourier basis, independent of the specific
profile of the periodic pattern and is, for ideal and noiseless signals,
related to the classic definition of contrast visibility using minimal
and maximal intensity values $(I_{\min}(r)-I_{\max}(r))/(I_{\min}(r)+I_{\max}(r))$
by a constant factor. This factor does depend on the specific profile
shape, and obviously is 1 for sinusoid profiles. For a square wave
of 50\% duty cycle, the actual contrast is $4/\pi\approx1.27$ times
higher than that of its first harmonic, as can be directly inferred
from its analytic Fourier transform. With respect to the explicit
contrast definition based on extremal values, Fourier analysis provides
the advantage of reduced noise susceptibility due to consideration
of all available data points as opposed to only the minimum and maximum
intensities. 

\subsection{Inference of modulation transfer, point spread and resolution}

When assuming a Gaussian point spread function $g(d)$ (of standard
deviation $\sigma$) with respect to a radial distance $d$ from the
optical path:
\begin{equation}
\hphantom{\mathrm{MTF}(f)}\mathllap{g(d)}=\mathrlap{\frac{1}{\sqrt{2\pi}\sigma}e^{-\frac{1}{2}\frac{d^{2}}{\sigma^{2}}}\approx\frac{0.9394}{\mathrm{FWHM}}e^{-\frac{2.773}{\mathrm{FWHM}^{2}}d^{2}}\:,}\hphantom{e^{-\frac{1}{2}(2\pi f\sigma)^{2}}\approx e^{-3.56\,\mathrm{FWHM}^{2}f^{2}}}\label{eq:PSF}
\end{equation}
the amplitude of a sinusoid function of period length $T$ or frequency
$f=T^{-1}$ is reduced by 
\begin{equation}
\mathrm{MTF}(f)=e^{-\frac{1}{2}(2\pi f\sigma)^{2}}\approx e^{-3.56\,\mathrm{FWHM}^{2}f^{2}}\label{eq:mtf}
\end{equation}
based on the Fourier transform of $g(d)$. FWHM thereby denotes the
full width at half maximum:
\begin{equation}
\hphantom{\mathrm{MTF}(f)}\mathllap{\mathrm{FWHM}(\sigma)}=\mathrlap{\sqrt{8\ln(2)}\,\sigma\approx2.355\,\sigma\:,}\hphantom{e^{-\frac{1}{2}(2\pi f\sigma)^{2}}\approx e^{-3.56\,\mathrm{FWHM}^{2}f^{2}}}
\end{equation}
which provides a more intuitive measure of point spread width as compared
to $\sigma$.

As the zeroth (mean) component of a signal will be unaffected by normalized
point spread functions, the above relation (Eq.~\ref{eq:mtf}) equivalently
describes the frequency dependent reduction of contrast (defined as
the ratio of signal amplitude and mean). Due to the orthogonality
of the Fourier basis, this relation is independent of the existence
of further harmonics and thus independent of the actual shape of the
considered periodic profile.

The maximum resolvable frequency $f_{\max}$ is commonly defined as
\begin{align}
\mathrm{MTF}(f_{\max}) & =0.1\:,\\
\intertext{\text{which for a Gaussian MTF implies}}f_{\max} & =\smash[t]{\frac{2\sqrt{\ln(2)\ln(10)}}{\pi\,\mathrm{FWHM}}\approx\frac{0.8043}{\mathrm{FWHM}}\:.}\label{eq:f_max}
\end{align}

The linearity of the Fourier transform furthermore guarantees that
the modulation transfer function (MTF) of a point spread function
described by a linear combination of multiple Gaussians (sharing the
same center point) corresponds to an equivalent linear combination
of their individual MTFs. This property allows for a straight forward
characterization of point spread widths whenever an experimentally
obtained MTF is sufficiently well describable by a linear combination
of few Gaussian functions.

The so determined point spread widths are to be understood in relation
to the sample dimensions and are, without further considerations,
specific to the chosen sample position relative to source and detector.
In general, geometric magnification relations need to be accounted
for with respect to the deduction of actual focal spot sizes of the
source. However, due to the negligible source--object distance as
compared to the source--detector distance (ratio $\approx10^{-3}$)
in the present case, the observed point spreads can be directly understood
also in spatial units at the X-ray focus.

In contrast to the frequency dependent characterization of contrast
visibility of periodic patterns by means of Eq.~\ref{eq:mtf}, the
derivative of an edge spread function directly represents the point
spread function in real space (as opposed to its Fourier transform,
the MTF) when assuming that the employed edge itself represents a
sufficiently perfect step function. Analogous to the previous considerations,
it may be characterized by a linear combination of Gaussian PSFs as
given by Eq~\ref{eq:PSF}.

\section{Results}

\subsection{MTF analysis from star pattern}

\begin{figure}[p]
\begin{centering}
\includegraphics[width=0.87\textwidth]{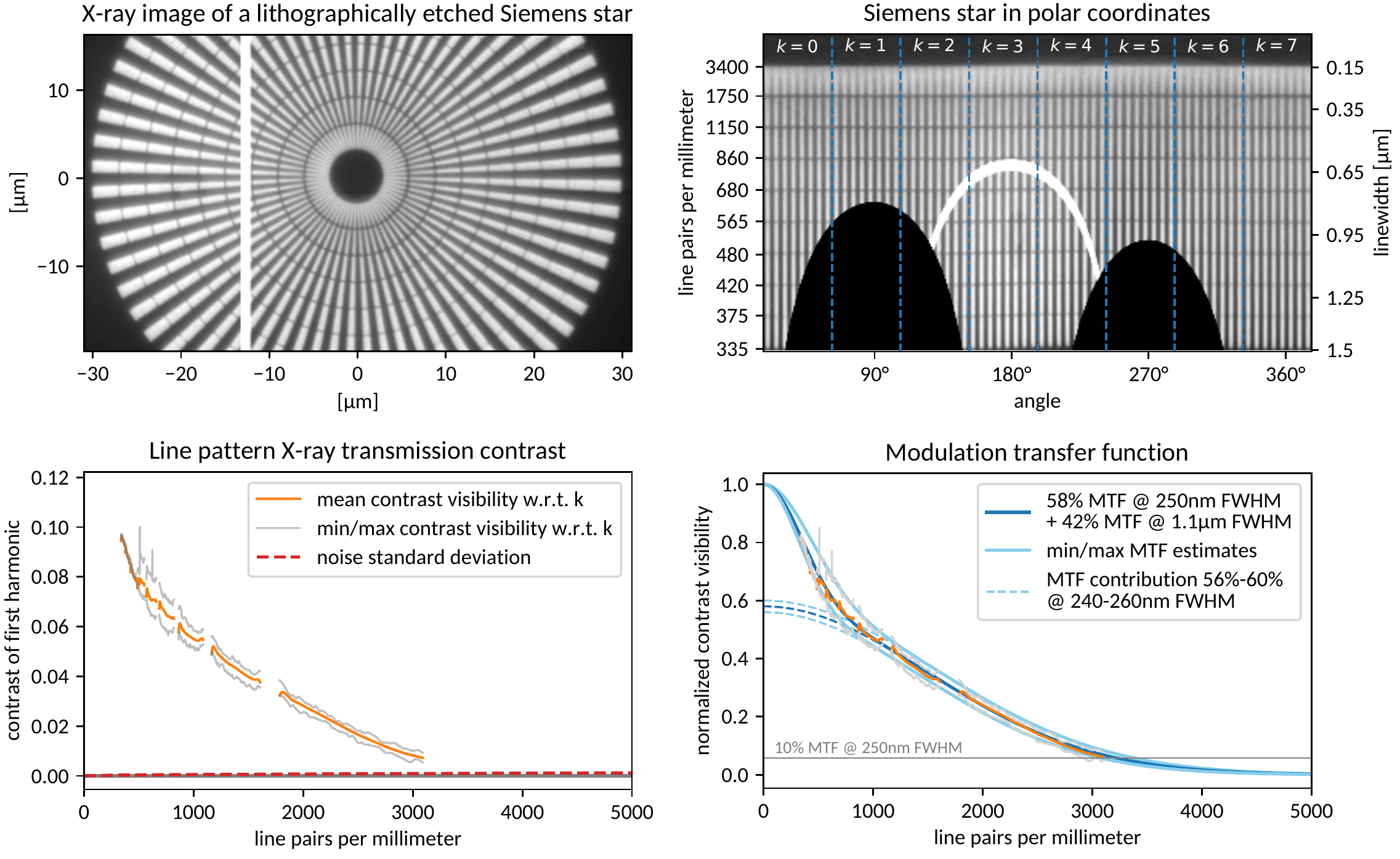}
\par\end{centering}
\caption{\label{fig:starmtf}X-ray transmission image of a tungsten based Siemens
star (upper left), its representation in polar coordinates (upper
right), quantification of frequency dependent transmission contrast
(lower left, cf.\ Eqs.~\ref{eq:visibility}--\ref{eq:mean-offset})
and the deduced modulation transfer function (lower right, cf.\ Eqs.~\ref{eq:mtf-min}--\ref{eq:mtf-max}).
The material is not fully absorbing, wherefore the observed contrast
of the first harmonic does not exceed 12\% (equivalent to 15\% contrast
for a square wave) in the low frequency limit. The grayscale window
of the shown images has therefore been visually adjusted to the object's
contrast. The X-ray image covers a subregion on the detector including
a band of 12 insensitive detector columns (white) in between detector
tiles, which transforms to a parabola in the polar representation.
Likewise, missing top and bottom sections transform to black parabolas.
Dashed lines in the polar representation separate individually analyzed
sections $k$ (cf.\ Section~\ref{subsec:siemens-star-analysis}).
The expected noise level of the contrast analysis is indicated in
red (lower left, cf.\ Eq.~\ref{eq:sigma_v})}
\end{figure}
\begin{figure}[p]
\begin{centering}
\includegraphics[width=0.87\textwidth]{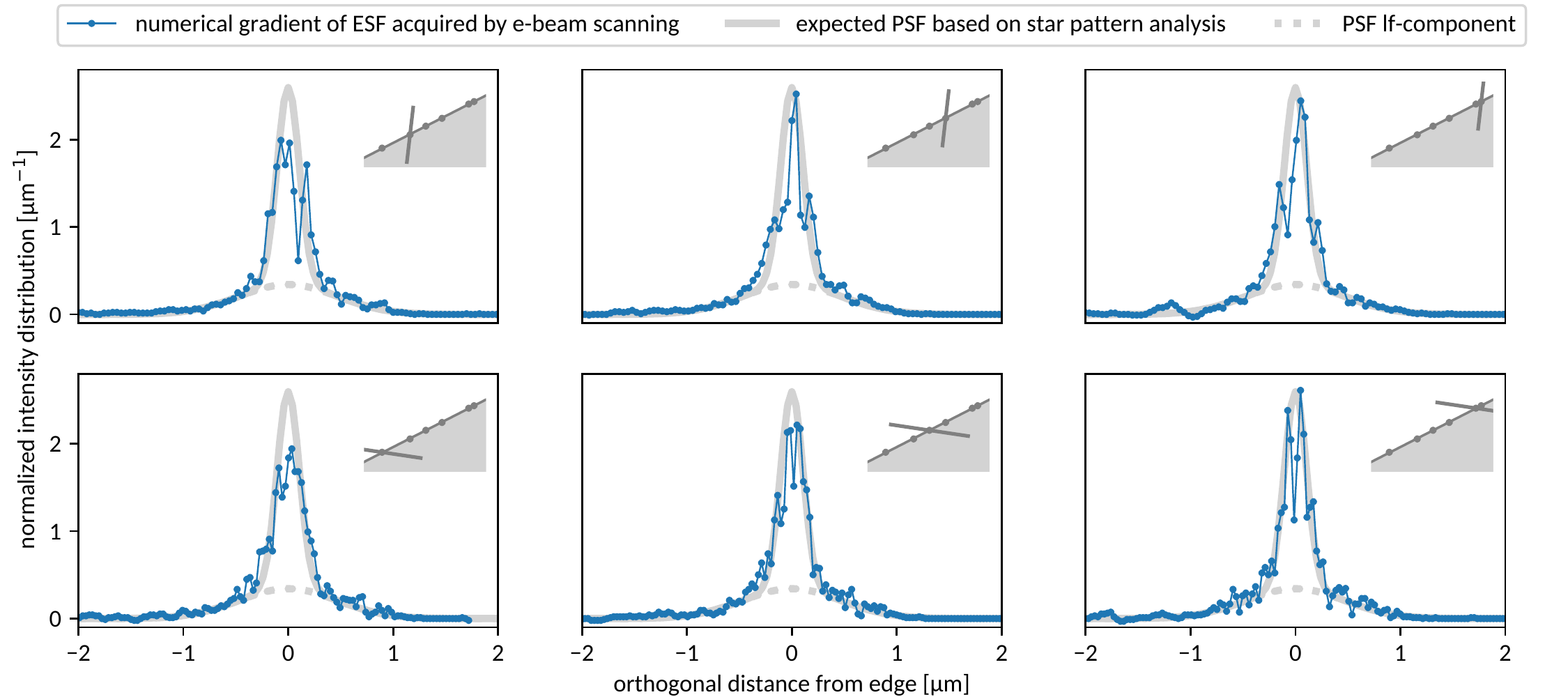}
\par\end{centering}
\caption{\label{fig:edge-scan-results}Point spread of the X-ray source's electron
beam determined by numerical differentiation of edge spread curves.
The latter are acquired by monitoring X-ray emission while scanning
the source's electron beam over an edge of its structured transmission
target. Scanning has been performed at shallow angles with respect
to the edge in order to increase the sampling density, and multiple
locations within a range of 9$\,$\textmu m along the edge have been
addressed. The respective scanning trajectories are indicated in the
upper right corner of each graph.  Although the edge scans are apparently
subject to considerable noise, they are in plausible agreement with
the point spread function -- indicated in gray -- derived previously
by imaging a star pattern (cf.\ Figure~\ref{fig:starmtf} and Eq.~\ref{eq:exp-PSF}).}
\end{figure}
Figure~\ref{fig:starmtf} (upper left) shows the X-ray projection
image of the Siemens star normalized to a flat field image acquired
without sample and scaled to the observed transmission contrast. 
The transmission of the tungsten background in the very center of
the pattern ranges at $t_{\min}\approx0.72$, with a statistical noise
of $\text{\ensuremath{\sigma_{I}}}\approx0.003$. Given the pixel
size of $p\approx0.07\,\text{\textmu m}$, smallest pattern period
of $T_{\min}=0.3\text{\textmu m}$, and $N/M=8$ (cf.\ Section~\ref{subsec:siemens-star-analysis}),
the uncertainty of the contrast visibility is found (using Eq.~\ref{eq:sigma_v})
to be smaller than
\begin{align}
\sigma_{v} & \lesssim0.001\:,
\end{align}
confirming the signal-to-noise ratio to be sufficient for the observed
range of contrast visibilities $v\gtrsim0.005$. 

In Figure~\ref{fig:starmtf} (upper right), the polar representation
of the image is given and annotated with the respective pattern frequencies
and line widths ($T/2$) along the radial dimension (cf.\ Eqs.\
\ref{eq:pattern-period} and \ref{eq:pattern-frequency}). The boundaries
between the eight angular segments indexed by $k$ and analyzed independently
by Eqs.~\ref{eq:visibility}--\ref{eq:mean-offset} are indicated
with vertical dashed lines. 

The bottom left panel of Fig.~\ref{fig:starmtf} shows the deduced
mean, minimum and maximum (with respect to $k$) contrast visibilities
as a function of pattern frequency. On the bottom right, linear combinations
of two Gaussians (shown in different shades of blue) characterizing
the high- and low frequency behaviors respectively have been manually
fitted to the mean, minimum and maximum modulation transfer characteristic
found within the Siemens star. The high frequency component is additionally
indicated separately using dashed lines. The frequency dependent contrast
reduction (MTF) within the star pattern is apparently well described
by this bi-Gaussian model (cf.\ Fig~\ref{fig:starmtf}), with
\begin{align}
\mathrm{MTF}_{\min}(f) & \approx0.56\,e^{-3.56\,(260\,\mathrm{nm})^{2}f^{2}}+0.44\,e^{-3.56\,(1.2\,\text{\textmu m})^{2}f^{2}}\label{eq:mtf-min}\\
\mathrm{MTF}_{\mathrm{mean}}(f) & \approx0.58\,e^{-3.56\,(250\,\mathrm{nm})^{2}f^{2}}+0.42\,e^{-3.56\,(1.1\,\text{\textmu m})^{2}f^{2}}\label{eq:mtf-mean}\\
\mathrm{MTF}_{\max}(f) & \approx0.60\,e^{-3.56\,(240\,\mathrm{nm})^{2}f^{2}}+0.40\,e^{-3.56\,(0.9\,\text{\textmu m})^{2}f^{2}}\:.\label{eq:mtf-max}\\
\intertext{\text{The point spread function corresponding to \ensuremath{\mathrm{MTF_{\mathrm{mean}}}}\ensuremath{(f)} is given by (cf. Eq. \ref{eq:PSF}):}}g_{\,\substack{\mathrm{mean}\\
\mathrm{MTF}
}
}(d) & \approx2.1794\,\text{\textmu m}^{-1}e^{-\frac{2.773\,d^{2}}{(250\,\mathrm{nm})^{2}}}+0.3587\,\text{\textmu m}^{-1}e^{-\frac{2.773\,d^{2}}{(1.1\,\text{\textmu m})^{2}}}\:,\label{eq:exp-PSF}
\end{align}
with full widths at half maximum 
\begin{align}
\mathrm{FWHM}_{(\mathrm{MTF},\mathrm{hf})} & \approx(250\pm10)\,\mathrm{nm}\\
\mathrm{FWHM}_{(\mathrm{MTF},\mathrm{lf})} & \approx(1.1\pm0.2)\,\text{\textmu m}
\end{align}
of the high frequency (hf) and low frequency (lf) contributions respectively,
as determined from the MTF. The wide tail of the point spread contributes,
based on the above model fits, $(42\pm2)\,\%$ of the integral amount
of light. 

The expected maximal resolution as defined by Eq.~\ref{eq:f_max}
ranges at 
\begin{align}
f_{\max} & \approx\frac{0.8043}{\mathrm{(250\pm10)\,\mathrm{nm}}}\approx(3220\pm130)\,\mathrm{linepairs}/\mathrm{mm}\\
\intertext{\text{corresponding to a minimal resolvable linewidth (at 50\% duty cycle) of}}\frac{T_{\min}}{2} & \approx\frac{1}{2}\frac{\mathrm{(250\pm10)\,\mathrm{nm}}}{0.8043}\approx(155\pm6)\,\mathrm{nm}\,
\end{align}
Due to the longer tailed component of the point spread function as
expected from the bi-Gaussian MTF, the actual contrast transfer at
$f_{\max}$ is expected to range at $(5.8\pm0.1)\,\%$.

\subsection{PSF from edge scan of the electron focus}

Figure~\ref{fig:edge-scan-results} compares Eq.~\ref{eq:exp-PSF}
to numerically differentiated edge scans of the source's electron
focus. Due to the considerable noise level observed, a total of 6
scans distributed over multiple positions along the edge have been
performed (covering a total range of $9\,\text{\textmu m}$ along
the edge). Two different scanning directions corresponding to the
two available deflection coils have been chosen. The actual scanning
motions have been calibrated to spatial units as outlined in Figure~\ref{fig:spatial-calibration}.
The true orientation of the scanned edge has subsequently been determined
as a line fit to the transition points of the acquired edge spread
functions. The transition points, edge and scanning trajectories are
explicitly indicated in Figure~\ref{fig:edge-scan-results} along
with the observed point spread functions.  

Both the narrow and wide components of the PSF as determined previously
from the X-ray image of a star pattern are visibly reproduced. Quantitative
least squares fits (not shown) of the bi-Gaussian PSF model are however
found to exhibit large variations of up to 30\% among repeated experiments
with respect to all fit parameters, indicating that the edge scans
are predominantly suited for qualitative assessments. Given the considered
length scales, the pronounced and reproducible noise right at the
very edge (i.e., at the very center of the deduced PSF) may have multiple
explanations, including in particular actual wear of the tungsten
edge due to frequent electron bombardment. 

Vanishing in the derivatives shown in Figure~\ref{fig:edge-scan-results}
is a constant offset within the edge spread functions corresponding
to 20\% of the total intensity. The respective point spread FWHM of
this contribution is inaccessible to the present methodologies and
can only be constrained to be either among the observed ones or range
above 2$\,$\textmu m based on the unobserved frequencies of the MTF
below 340 line pairs per millimeter.

\section{Discussion}

The analysis of the Siemens star provides experimental data on the
frequency dependent contrast reduction in the range between 340 and
3100 line pairs per millimeter. Although the finest structures within
the imaged star actually range up to about 3400 line pairs per millimeter,
regions at the very end of each patterned band within the star have
been intentionally excluded from the analysis due to systematic contrast
enhancements caused by radial discontinuities of the bar pattern.
The observed contrasts can be well described by a bi-Gaussian (yet
single-peak) model of the MTF and the corresponding PSF, which reveals
point spread FWHMs of $(250\pm10)\,\mathrm{nm}$ and $(1.1\pm0.2)\,\text{\textmu m}$
for the narrow and wide components of the PSF respectively. The narrow
PSF component determining the maximal achievable imaging resolution
thereby contributes about half of the total X-ray emission, and further
ranges at the better end of expectations from Monte Carlo simulations
of X-ray generation in thin target layers \cite{Nachtrab2011}. These
results are further supported by an independent assessment of the
X-ray source's electron focal spot obtained by scanning the electron
beam over an edge of the source's transmission target and analyzing
the variation in X-ray emission. Despite being less suited for quantitative
analyses (in accordance with the initial motivation for the use of
periodic patterns as opposed to edges), the direct observation of
point spread shows to be consistent with the observed modulation transfer
within the star pattern and in particular confirms the two-component
nature of the PSF. In addition, a contribution of 20\% of the total
emitted intensity can be observed when focusing the electron beam
on the bare diamond vacuum window, which may originate from the vacuum
window itself, from de-focused electrons hitting distant target regions,
or from residual metals on the vacuum window. Bremsstrahlung generated
within the vacuum window (which also serves as the structured target's
substrate) is indeed expected from Monte Carlo simulations \cite{Zhou2016},
and would, due to the window's much larger thickness as compared to
the actual target material, indeed exhibit a considerably wider FWHM
as compared to the main contribution from the target layer.

\section{Conclusion}

The modulation transfer function (MTF) and respective point spread
function (PSF) characterizing the imaging performance of the ntCT
nano-tomography system and the employed Excillum NanoTube X-ray source
have been determined both by means of Fourier analysis of the image
of a Siemens star and edge scans of the focused electron beam over
the source's structured tungsten target. Within the frequency range
of 340 to 3100 line pairs per millimeter a bi-Gaussian (yet single-peak)
point spread shape is found, whereof the narrow contribution exhibits
a FWHM of $(250\pm10)\,\mathrm{nm}$, allowing for imaging applications
with resolutions up to $150\,\mathrm{nm}$.  About half of the total
X-ray intensity is found to be emitted from this narrow focal point.

\section{Appendix: noise analysis}

Particularly with respect to the highest frequencies at the resolution
limit, quantification of the signal-to-noise ratio of the present
contrast visibility analyses is critical for multiple reasons: First,
the analysis of (non-negative) amplitude magnitudes (Eq.~\ref{eq:amplitude})
is susceptible to non-negligible bias whenever the statistical noise
is comparable to or even exceeds the actual amplitudes (cf.\ e.g.\
\cite{Gudbjartsson1995}). Noise is here  expected to increase for
smaller structures due to the nature of the Siemens star, with smaller
structures also covering less detector area. Smaller structures are
finally expected to exhibit smaller amplitudes due to the imaging
system's modulation transfer properties. The expected noise variance
of $\sigma_{v}^{(k)}(r)$ of $v^{(k)}(r)$ with respect to the radial
distance $r$ from the center and the related pattern frequency $f(r)=1/T(r)$
(cf.\ Eqs.~\ref{eq:pattern-period}--\ref{eq:pattern-frequency})
shall thus be explicitly related to the directly quantifiable variance
$\sigma_{I}^{2}$ of the original image intensities.

First, the noise properties of $a_{1}^{(k)}(r)$ and $t^{(k)}(r)$
need to be considered. Given that the integrals of Eqs.~\ref{eq:amplitude}--\ref{eq:mean-offset}
will generally be evaluated as finite sums over discrete intensity
samples $I(r,\varphi_{i})$ at $i\in\{1,...,n\}$, the following holds
under the assumption of uncorrelated Gaussian noise and equidistant
phases $\varphi_{i}$ covering multiples of a full period of the considered
pattern:
\begin{align}
\sigma_{a_{1}}^{(k)}(r) & \approx\sum_{i}\sqrt{\left(\frac{\partial a_{1}^{(k)}}{\partial I(r,\varphi_{i})}\,\sigma_{I}\right)^{2}}=\sqrt{\frac{2}{n}}\,\sigma_{I}\\
\sigma_{t}^{(k)}(r) & \approx\sum_{i}\sqrt{\left(\frac{\partial a_{1}^{(k)}}{\partial I(r,\varphi_{i})}\,\sigma_{I}\right)^{2}}=\sqrt{\frac{1}{n}}\,\sigma_{I}\:,
\end{align}
with $n$ denoting the number of intensity samples (i.e., detector
pixels) contributing to the evaluation of $v^{(k)}(r)$ at a given
radius $r$ of the star pattern. The number of actual detector pixels
contributing can be straight forwardly estimated by comparing the
lengths of a considered bar pattern (i.e., $N/M$ periods of length
$T(r)=1/f(r)$) to the known pixel size $p$ of the original (prior
to upsampling) images:
\begin{align}
n & \approx\frac{N\,T(r)}{M\,p}\:.
\end{align}
Based on the orthogonality of the Fourier basis, the noise on $a_{1}^{(k)}(r)$
and $t^{(k)}(r)$ remains uncorrelated, such that
\begin{equation}
\sigma_{v}^{(k)}(r)\approx\sqrt{\left(\frac{\partial v^{(k)}}{\partial a_{1}^{(k)}(r)}\,\sigma_{a_{1}}^{(k)}(r)\right)^{2}+\left(\frac{\partial v^{(k)}}{\partial t(r)}\,\sigma_{t}^{(k)}(r)\right)^{2}}\approx\frac{\sigma_{a_{1}}^{(k)}(r)}{t^{(k)}(r)}\qquad\text{for }a_{1}^{(k)}(r)\ll t^{(k)}(r)
\end{equation}
holds analogously.

The largest noise variance on $v^{(k)}(T(r))$ is thus expected for
the smallest pattern period $T_{\min}$ (or highest frequency $f_{\max}=1/T_{\min}$)
within in the Siemens star:
\begin{align}
\sigma_{v}(f=1/T) & \lesssim\frac{1}{t_{\min}}\sqrt{2\frac{M}{N}\frac{p}{T_{\min}}}\sigma_{I}\:,\label{eq:sigma_v}
\end{align}
where $t_{\min}$ characterizes the minimal mean transmission value
observed within the star pattern. A lower bound may e.g.\ be determined
from the unstructured center region (cf.\ Figure~\ref{fig:starmtf}).
Given that the properties $\sigma_{I}$ (image noise standard deviation)
and $t_{\min}$ (lower bound on the mean transmission of the star
pattern) are representative for the entire image, the upper bound
on $\sigma_{v}$ given by Eq.~\ref{eq:sigma_v} is independent of
the considered angular section $k$.

\section*{Acknowledgements}

Funding is acknowledged from the Horizon 2020 project no.\ 814485.
R.\ Hanke and S.\ Zabler are acknowledged for securing funding.

\paragraph{Authors' contributions:}

J.G.\ conceived the methodology, did the data analysis and wrote
the manuscript. C.F., D.M., A.B.\ constructed the ntCT, performed
the experiments and critically revised the manuscript.

\bibliographystyle{plain}

\end{document}